\documentclass[%
 aip,
 amsmath,amssymb,
 reprint,%
]{revtex4-1}

\usepackage{graphicx}% Include figure files
\usepackage{dcolumn}% Align table columns on decimal point
\usepackage{bm}% bold math
\usepackage[utf8]{inputenc}
\usepackage[T1]{fontenc}
\usepackage{mathptmx}
\usepackage{array}
\usepackage{etoolbox}
\usepackage{multirow}
\usepackage{makecell}
\usepackage{subcaption}
\usepackage{xcolor}

\makeatletter
\def\@email#1#2{%
 \endgroup
 \patchcmd{\titleblock@produce}
  {\frontmatter@RRAPformat}
  {\frontmatter@RRAPformat{\produce@RRAP{*#1\href{mailto:#2}{#2}}}\frontmatter@RRAPformat}
  {}{}
}%
\makeatother
\begin{document}

\preprint{AIP/123-QED}

\title{Updates on MURAVES Project at Mt. Vesuvius}
% Force line breaks with \\

\author{Yanwen Hong}%$^{*,9,10}$,
 \email{yanwen.hong@cern.ch}
\affiliation{Inter-University Institute for High Energies, Vrije Universiteit Brussel, Belgium}
\affiliation{Department of Physics and Astronomy, Ghent University, Belgium}

\author{Marwa Al Moussawi}%$^{1}$,
\altaffiliation[Now at ]{Muon vision, Inc, Cambridge MA, USA}
\affiliation{Centre for Cosmology, Particle Physics and Phenomenology, Université catholique de Louvain, Belgium}

\author{Fabio Ambrosino}%$^{2,3}$, 
\affiliation{National Institute of Nuclear Physics - Section of Naples, Italy}
\affiliation{Department of Physics, University of Naples Federico II, Italy}

\author{Antonio Anastasio}%$^{2}$,
\affiliation{National Institute of Nuclear Physics - Section of Naples, Italy}

\author{Samip Basnet}
\altaffiliation[Now at ]{Tsung-Dao Lee Institute and School of Physics and Astronomy, Shanghai Jiao Tong University, Shanghai 201210, China}
\affiliation{Centre for Cosmology, Particle Physics and Phenomenology, Université catholique de Louvain, Belgium}

\author{Lorenzo Bonechi}%$^{4}$,
\affiliation{National Institute of Nuclear Physics - Section of Florence, Italy}

\author{Diletta Borselli}%$^{4,6}$,
\affiliation{National Institute of Nuclear Physics - Section of Florence, Italy}
\affiliation{Department of Physics and Geology, University of Perugia, Italy}

\author{Alan Bross}%$^{7}$,
\affiliation{Fermi National Accelerator Laboratory, USA}

\author{Antonio Caputo}%$^{8}$,
\affiliation{National Institute of Geophysics and Volcanology - Vesuvius Observatory, Italy}

\author{Roberto Ciaranfi}%$^{4}$,
\affiliation{National Institute of Nuclear Physics - Section of Florence, Italy}

\author{Luigi Cimmino}%$^{2,3}$, 
\affiliation{National Institute of Nuclear Physics - Section of Naples, Italy}
\affiliation{Department of Physics, University of Naples Federico II, Italy}

\author{Vitaliano Ciulli}%$^{4,5}$,
\affiliation{National Institute of Nuclear Physics - Section of Florence, Italy}
\affiliation{Department of Physics, University of Florence, Italy}

\author{Raffaello D'Alessandro}%$^{5}$,
\affiliation{Department of Physics, University of Florence, Italy}

\author{Catalin Frosin}%$^{4,5}$,
\affiliation{National Institute of Nuclear Physics - Section of Florence, Italy}
\affiliation{Department of Physics, University of Florence, Italy}

\author{Gabor Nyitrai}%$^{2,3}$,
\affiliation{National Institute of Nuclear Physics - Section of Naples, Italy}
\affiliation{Department of Physics, University of Naples Federico II, Italy}

\author{Andrea Giammanco}%$^{1}$,
\affiliation{Centre for Cosmology, Particle Physics and Phenomenology, Université catholique de Louvain, Belgium}

\author{Flora Giudicepietro}%$^{8}$,
\affiliation{National Institute of Geophysics and Volcanology - Vesuvius Observatory, Italy}

\author{Sandro Gonzi}%$^{4,5}$, 
\affiliation{National Institute of Nuclear Physics - Section of Florence, Italy}
\affiliation{Department of Physics, University of Florence, Italy}

\author{Giovanni Macedonio}%$^{8}$, 
\affiliation{National Institute of Geophysics and Volcanology - Vesuvius Observatory, Italy}

\author{Vincenzo Masone}%$^{2}$, 
\affiliation{National Institute of Nuclear Physics - Section of Naples, Italy}

\author{Massimo Orazi}%$^{8}$,
\affiliation{National Institute of Geophysics and Volcanology - Vesuvius Observatory, Italy}

\author{Andrea Paccagnella}%$^{4,5}$,
\affiliation{National Institute of Nuclear Physics - Section of Florence, Italy}
\affiliation{Department of Physics, University of Florence, Italy}

\author{Rosario Peluso}%$^{8}$,
\affiliation{National Institute of Geophysics and Volcanology - Vesuvius Observatory, Italy}

\author{Anna Pla-Dalmau}%$^{7}$,
\affiliation{Fermi National Accelerator Laboratory, USA}

\author{Amrutha Samalan}%$^{9}$,
\altaffiliation[Now at ]{Paul Scherrer Institute, Forschungsstrasse 111, 5232 Villigen PSI, Switzerland}
\affiliation{Department of Physics and Astronomy, Ghent University, Belgium}

\author{Giulio Saracino}%$^{3}$,
\affiliation{National Institute of Nuclear Physics - Section of Naples, Italy}
\affiliation{Department of Physics, University of Naples Federico II, Italy}

\author{Giovanni Scarpato}%$^{8}$,
\affiliation{National Institute of Geophysics and Volcanology - Vesuvius Observatory, Italy}

\author{Paolo Strolin}%^{2,3}$, 
\affiliation{National Institute of Nuclear Physics - Section of Naples, Italy}
\affiliation{Department of Physics, University of Naples Federico II, Italy}

\author{Michael Tytgat}%$^{9,10}$,
\affiliation{Inter-University Institute for High Energies, Vrije Universiteit Brussel, Belgium}
\affiliation{Department of Physics and Astronomy, Ghent University, Belgium}

\author{Enrico Vertechi}%$^{8}$ }
\affiliation{National Institute of Geophysics and Volcanology - Vesuvius Observatory, Italy}

\date{\today}% It is always \today, today,

\begin{abstract}
The MUon RAdiography of VESuvius (MURAVES) project aims to use muography imaging techniques to study the
internal structure of the summit of the Mt. Vesuvius, an active volcano near Naples, Italy. This paper presents recent advancements in both data analysis and simulation tools that enhance the quality and reliability of the experiment’s results. A new track selection method, termed the Golden Selection, has been introduced to select high-quality muon tracks by applying a refined $\chi^2$-based criterion. This selection improves the signal-to-background ratio and enhances the resolution of muographic images. Additionally, the simulation framework has been upgraded with the integration of the MULDER (MUon simuLation for DEnsity Reconstruction) library, which unifies the functionalities of pervious used libraries within a single platform. MULDER enables efficient and accurate modeling of muon flux variations due to topographical features. An agreement is shown between simulated and experimental flux map.

\end{abstract}

\maketitle

\section{Introduction}

Mount Vesuvius, located near Naples, Italy, is one of the most hazardous volcanoes due to its history of explosive eruptions and the dense population in its vicinity. Understanding its internal structure is crucial for assessing potential eruption scenarios and implementing effective risk mitigation strategies~\cite{de2020invited}. Traditional geophysical methods have provided valuable insights but often face limitations in resolution and depth penetration~\cite{linde20173}.

Muon radiography, or muography, offers a novel approach by exploiting the natural flux of cosmic-ray muons to probe the internal density distribution of large structures like volcanoes. Muons are highly penetrating particles, making them suitable for imaging geological formations~\cite{tanaka2023muography}.

The MURAVES project \cite{d2019volcanoes,saracino2017muraves} aims to apply the muography technique to Mt. Vesuvius to image the internal structure of the summit crater. 
The experimental setup, located inside a green-coloured container on the slope of the crater approximately 1.5 km from the summit, consists of three identical muon trackers (ROSSO, NERO, and BLU), each covering a sensitive area of 1~$m^{2}$. Every tracker is composed of four XY tracking planes made from plastic scintillators. The X and Y tracking layers are built using 64 plastic scintillator bars per layer, with an isosceles triangular cross-section measuring 3.3 cm at the base and 1.7 cm in height. This design ensures that each muon passes through at least two adjacent bars, allowing for precise position reconstruction. The scintillation light produced by muon interactions is collected via wavelength-shifting (WLS) fibers embedded in the scintillator bars. These fibers guide the light to silicon photomultipliers (SiPMs) for detection. To suppress background from low-energy muons and other particles, a 60~cm thick lead wall is placed between the third and fourth tracking planes in each tracker. This passive shielding increases the energy threshold for detectable muons, effectively filtering out low-energy particles that are more susceptible to multiple scattering and energy loss, which can degrade image resolution. The experiment is currently operational, continuously collecting data from atmospheric cosmic-ray muons.

The following two chapters present the recent developments in the MURAVES analysis chain. Chapter2 introduces the \textit{Golden Selection} criteria for track reconstruction, a $\chi^2$-based method designed to enhance data quality of muon tracks. Chapter3 describes the integration of the new MULDER simulation framework, which unifies the modeling of muon transport through the volcanic structure. The results show improved agreement between experimental and simulated data.
\section{Data Analysis}
In the MURAVES experiment, muon track reconstruction is performed using four XY tracking planes within each detector module. As a muon passes through the detector, it produces scintillation light in the plastic scintillator bars, which is subsequently collected by the readout system. The resulting signals allow the determination of hit positions in both the X and Y coordinates across the four planes. By applying a linear fit to these four coordinate points, the muon's trajectory is accurately reconstructed, enabling precise analysis of its path through the detector. Further details on the reconstruction algorithm and the full data analysis chain are provided in~\cite{MariaelenaThesis}. Following the tracks reconstruction, a selection criterion based on track's $\chi^2$ value, as defined in Eq.~\ref{eq:chi2}, is applied to assess the quality of the fitted tracks. 
\begin{equation}
    \chi^2 = \sum^{n}_{i}\frac{(x_{i}^{model}-x_{i}^{data})^2}{\sigma^{2}_{x}}
    \label{eq:chi2}
\end{equation}

A lower $\chi^2$ implies a better alignment of the clusters, and is therefore useful for selecting the most accurate track when multiple tracks have been reconstructed in the same event. To this end, a cut on the $\chi^2$ value is applied to select a better sample of tracks, referred to as "golden" tracks—forming the basis of the so-called {\it Golden Selection}. This helps to enhance the quality of the muographic images and improve the accuracy of the density measurements~\cite{Samip}.

\begin{figure}[h!]
    \centering
    \begin{subfigure}[h!]{0.44\textwidth}
        \includegraphics[width=0.9\linewidth]{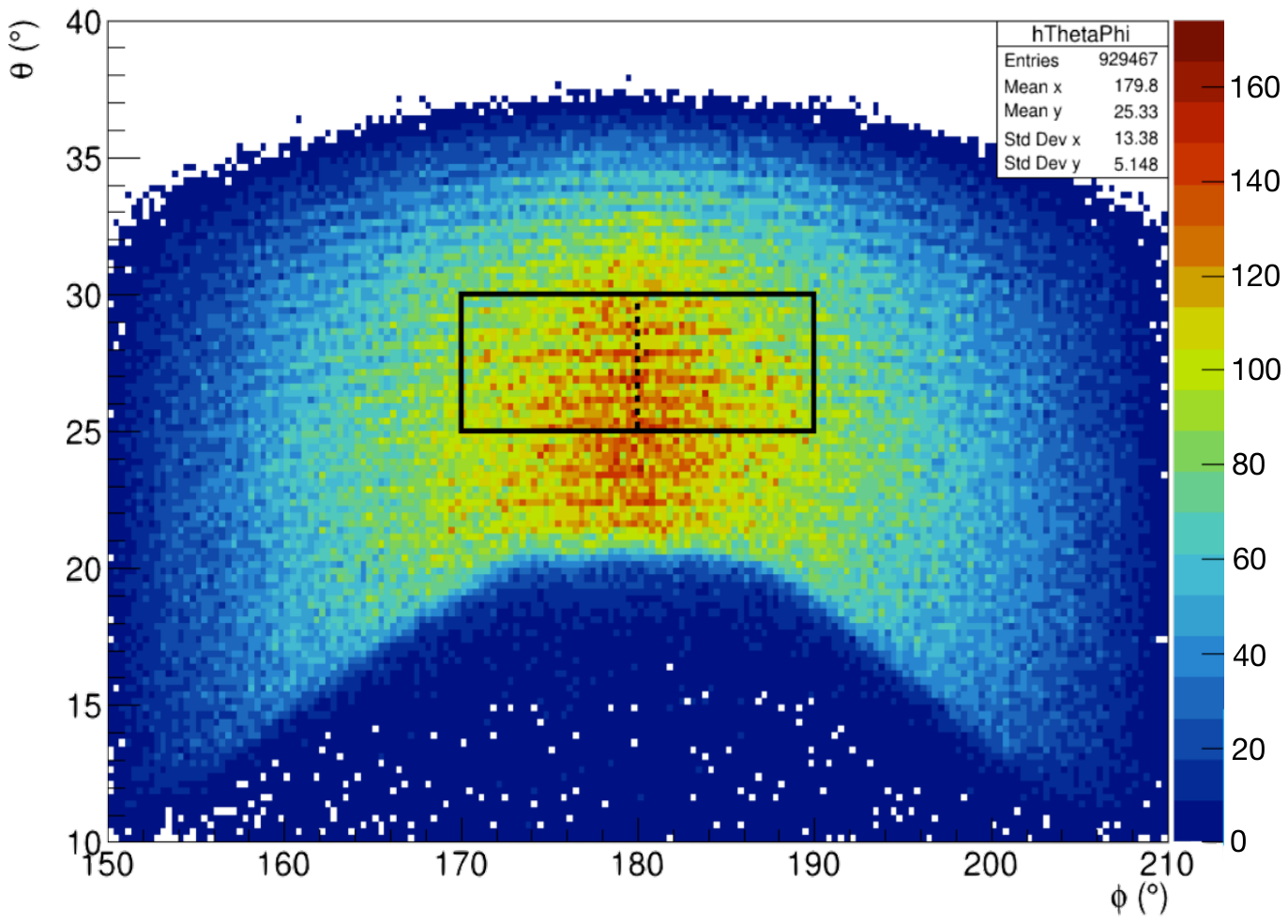}
        \caption{Illustration of control region highlighted in black box from Vesuvius datasets.}
        \label{fig:CRVes}
    \end{subfigure}
    \hfill
     \begin{subfigure}[h!]{0.44\textwidth}
        \includegraphics[width=0.9\linewidth]{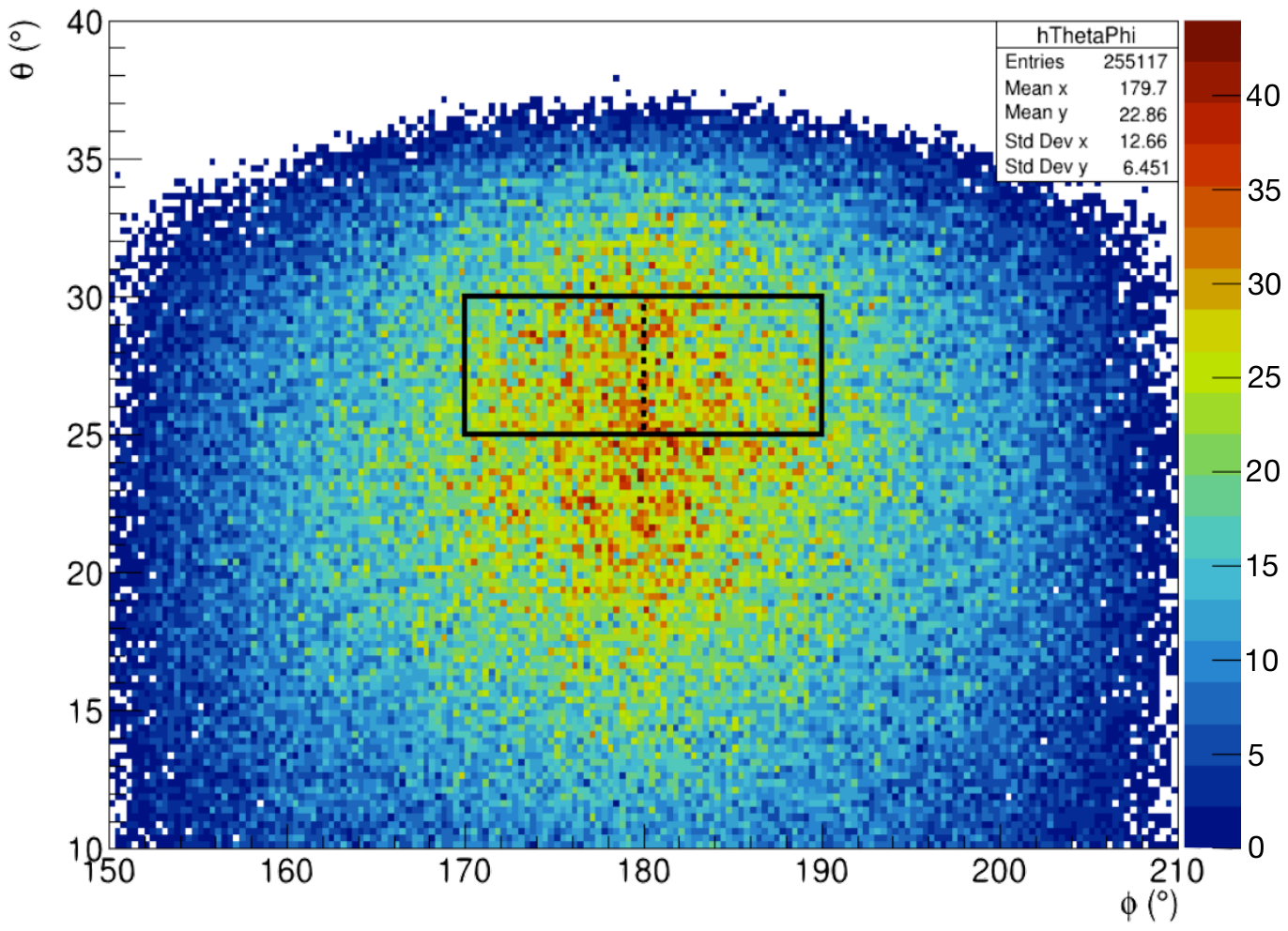}
         \caption{Illustration of control region highlighted in black box from free-sky datasets.}
         \label{fig:CRCal}
    \end{subfigure}
    \caption{Control regions shown in $\theta-\phi$ flux map of NERO detector at Working Point $15^{\circ}$.}
    \label{fig:CR}
\end{figure}

Since the cluster positions in the fourth plane are affected by the lead block between the third and the fourth stations, the $\chi^2$ used in the Golden Selection is computed using only the first three planes, denoted as $\chi^2_{3p}$. 
The current strategy relies on the assumption that the muon flux measured from open-sky directions above Mt. Vesuvius should be consistent with the flux measured during calibration runs—i.e., under free-sky conditions—along the corresponding directions. To validate this, a control region, shown in Fig. \ref{fig:CR}, is defined. This region is selected to lie sufficiently far from the volcano’s contours to avoid systematics arising from scattering or absorption at the edges of the structure. Within this control region, the following condition is expected to hold between the fluxes measured in the Vesuvius and calibration datasets:
\begin{equation}
    \frac{\Phi ^{Ves}_{\mu}(\Delta \theta, \Delta \phi)}{\Phi ^{Calib}_{\mu}(\Delta \theta, \Delta \phi)} =1 
    \label{eq:condition}
\end{equation}

\begin{figure}[h!]
    \centering
    \includegraphics[width=0.6\linewidth]{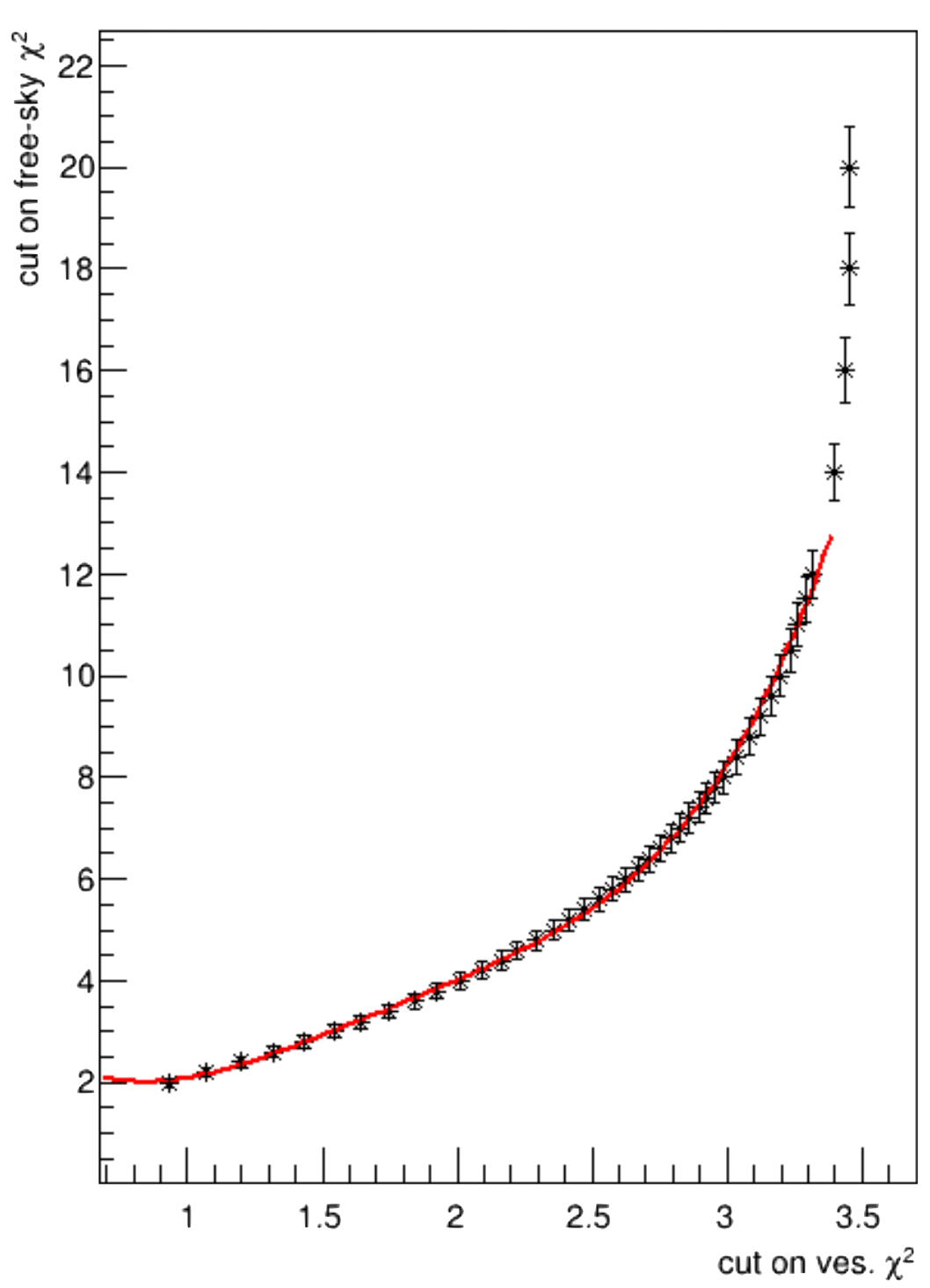}
    \caption{The $\chi^2_{3p}$ combination on the Vesuvius and the free-sky datasets satisfying the $T (\theta, \phi) = 1$ condition in the control region.}
    \label{fig:cut1}
\end{figure}

Based on the condition defined in Eq. \ref{eq:condition}, independent $\chi^2_{3p}$ cuts are applied to both the Vesuvius and free-sky datasets. The combination of these cuts, i.e., the pair of $\chi^2_{3p}$ thresholds for the Vesuvius and free-sky data that satisfy the condition, is illustrated in Fig. \ref{fig:cut1}.

\begin{figure}[h!]
    \centering
        \includegraphics[width=0.95\linewidth]{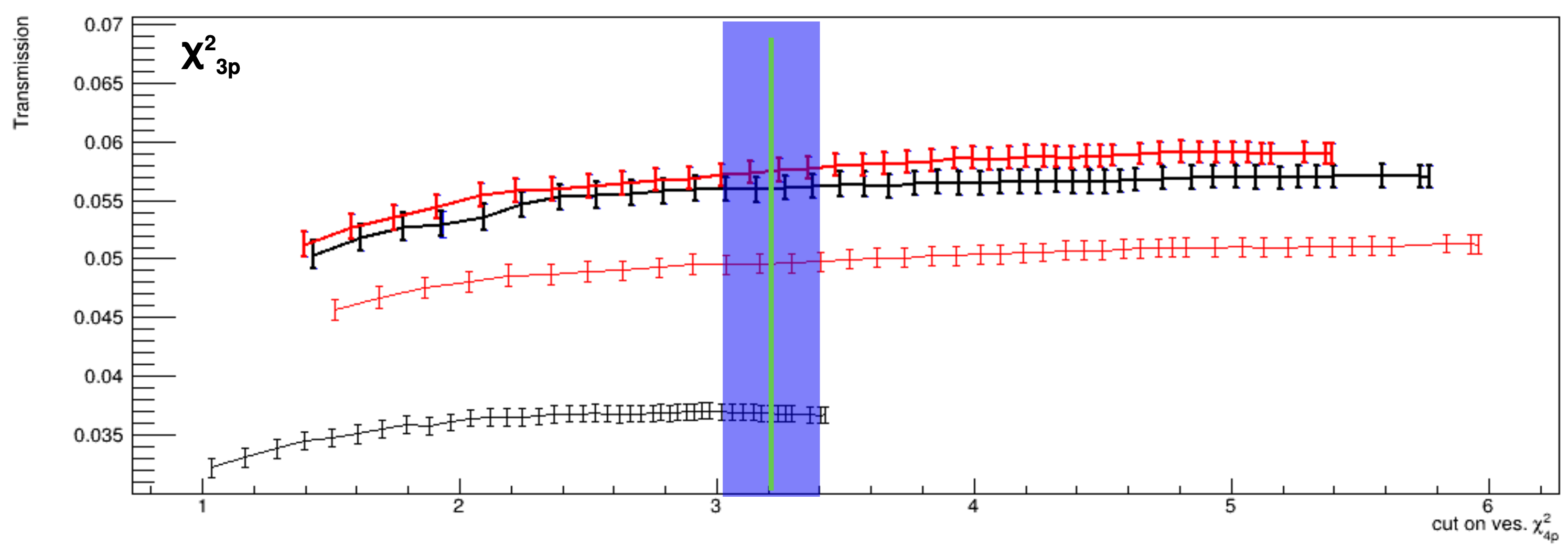}
        \caption{Measured transmission as a function of the maximum $\chi^2_{3p}$ cut value in the Vesuvius region. The black and red curves represent NERO and ROSSO datasets, with bold representing WP20 and normal, WP15.}
        \label{fig:cut3}
\end{figure}

The selected combination of $\chi^2_{3p}$ cuts is now used to evaluate the time-normalized muon transmission through the \textit{Vesuvius Region}, defined by the angular intervals $\phi \in [170^\circ, 190^\circ]$ and $\theta \in [12^\circ, 18^\circ]$. The effective ratio between the measured and calibration fluxes, referred to as the transmission $T(\theta, \phi)$, accounts for and largely cancels out detector-related effects. It is calculated using the following expression:
\begin{equation}
    T (\theta, \phi) =C \cdot \frac{N_{mes}(\theta, \phi)}{N_{cal}(\theta, \phi)}
\end{equation}
\noindent where $N_{mes}(\theta, \phi)$ and $N_{cal}(\theta, \phi)$ represent the angular distributions of muon counts from the Vesuvius and calibration datasets, respectively, and $C$ is a normalization constant. The measured transmission in the Vesuvius region for all datasets is shown in Fig. \ref{fig:cut3}, as a function of $\chi^2_{3p}$ threshold on Vesuvius datasets. A common upper cut on $\chi^2_{3p}$ was applied across all detectors and working points (WP), set to a value of 3.2 (indicated by the green line in the figure). This threshold was chosen based on the plateau region highlighted by the blue band, which indicates stability in the transmission values.

\begin{figure}[h!]
    \centering
        \includegraphics[width=0.75\linewidth]{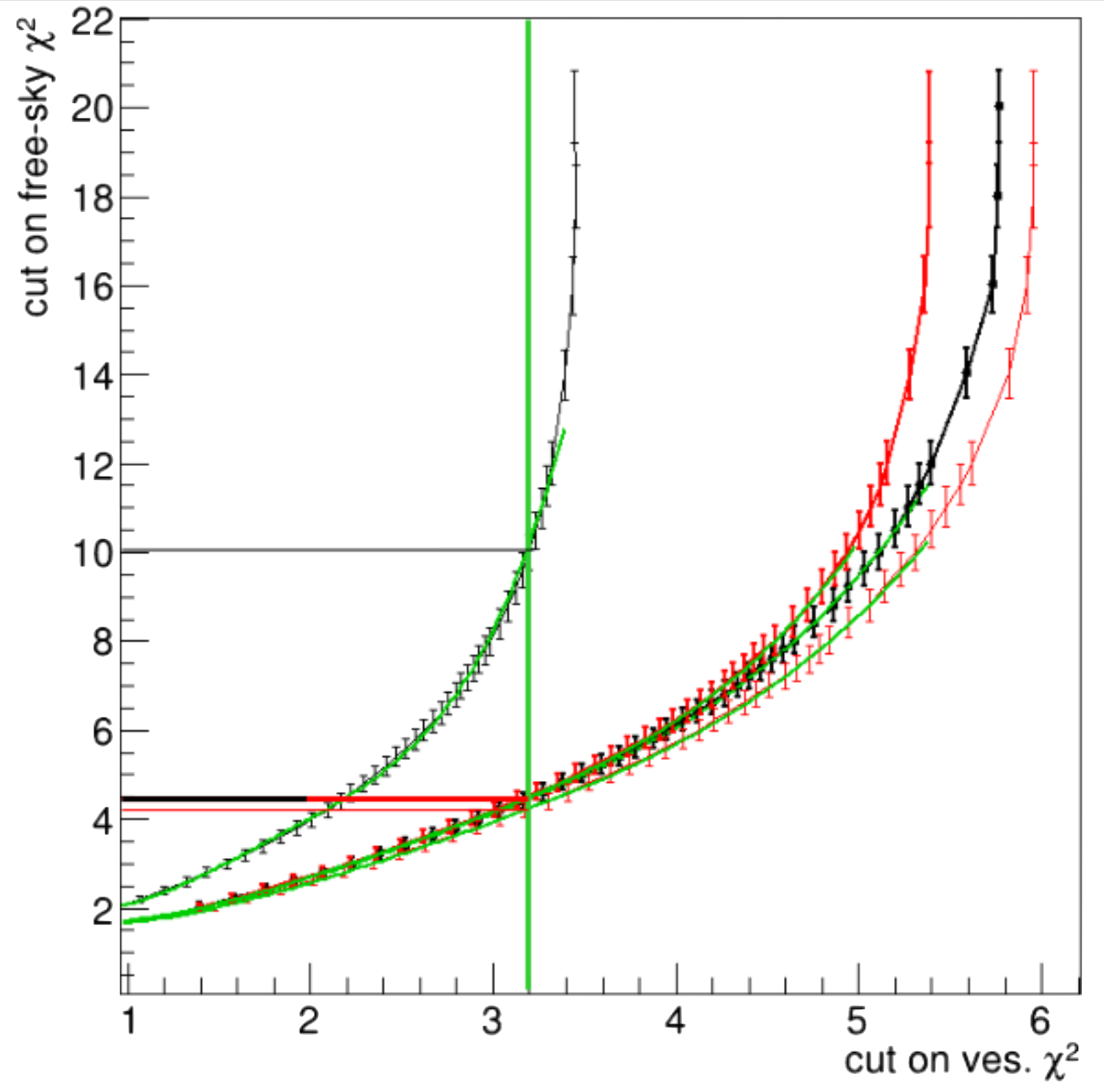}
        \caption{The $\chi^2_{3p}$ combination on the Vesuvius and the free-sky datasets satisfying the condition defined in Eq. \ref{eq:condition} for all available datasets. The black and red curves represent NERO and ROSSO datasets, with bold and normal representing WP20 and WP15, respectively.}
        \label{fig:cut4}
\end{figure}

However, due to differences between the Vesuvius and calibration datasets, including acquisition periods, environmental conditions, and detector operating states, the optimal free-sky $\chi^2_{3p}$ thresholds are not identical across all datasets. Based on the fixed cut value of 3.2 applied to the Vesuvius datasets, the corresponding optimal $\chi^2_{3p}$ cut values for the free-sky datasets were determined from the cut combination analysis shown in Fig. \ref{fig:cut4}. The resulting values are: 10.0 for NERO WP15, 4.5 for NERO WP20, 4.3 for ROSSO WP15, and 4.5 for ROSSO WP20.

\begin{figure}[h!]
    \centering
        \includegraphics[width=0.8\linewidth]{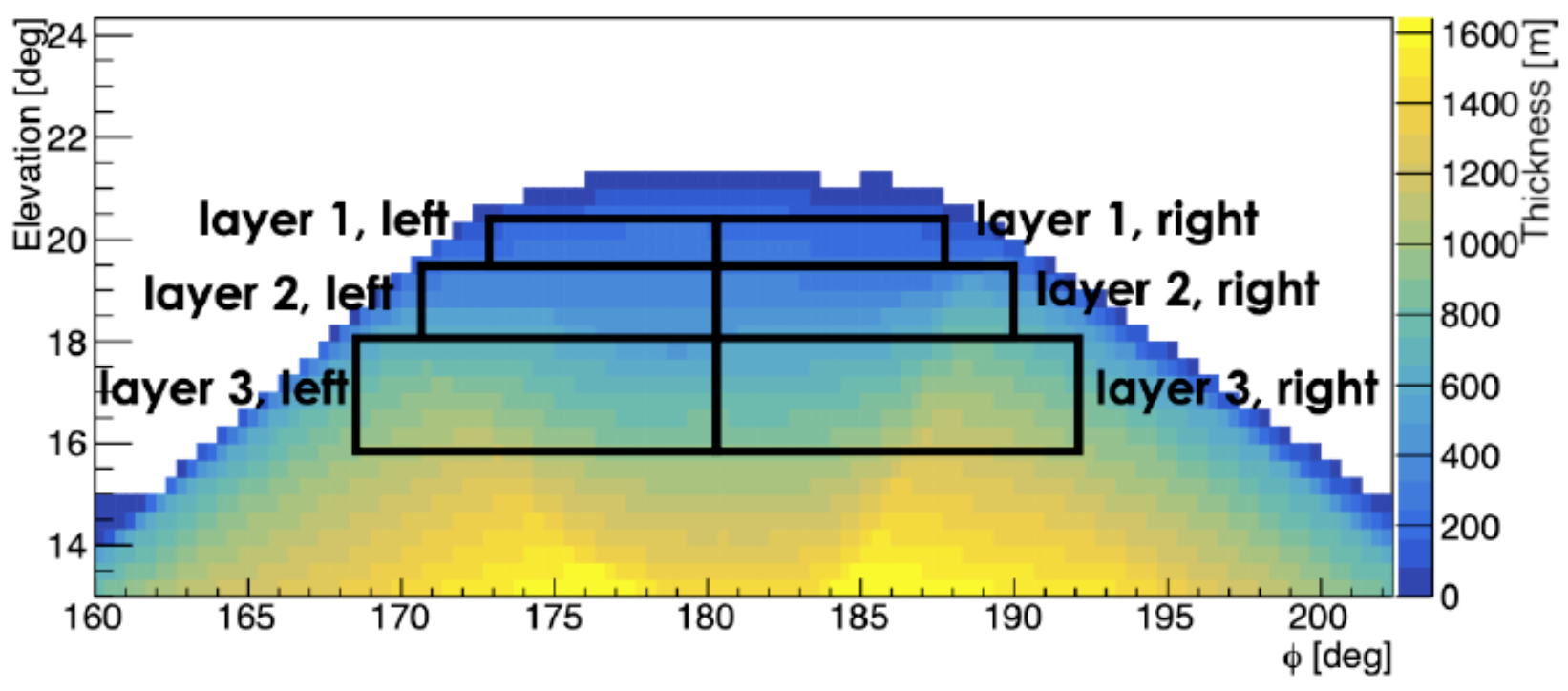}
        \caption{The rock thickness map of Mt. Vesuvius summit crater observed in MURAVES experiment location, and the angular regions (in bold black boxes) used for the study.}
        \label{fig:region}
\end{figure}

In muography applications to volcanology, the muon flux is significantly attenuated when traversing rock volumes with thicknesses on the order of several thousand meters. Therefore, in our study the region of interest is focused on the summit cone of Mt. Vesuvius, where the typical rock thickness is less than 1200 meters. The $\theta$–$\phi$ angular regions selected for this study are shown in Fig. \ref{fig:region}. These regions span three layers of elevation angles chosen across the summit cone and are divided into left and right azimuthal sectors to enable the investigation of possible left-right asymmetries. The specific angular intervals used in the analysis are detailed in Table \ref{ta:region}.

\begin{table}[h!]
\centering
\begin{tabular}{m{3.8cm} | m{2.22cm} | m{2.2cm}}
     Elevation $\Delta \theta$ & Left ($\Delta \phi$) & Right ($\Delta \phi$) \\
     \hline
      $\text{Top}\; _{\text{(layer1)}}$ $\quad \;$ [19.5$^\circ$ - 20.5$^\circ$]  & [173.0$^\circ$ - 180.0$^\circ$] & [180.0$^\circ$ - 187.0$^\circ$] \\
     \hline
     $\text{Middle}\; _{\text{(layer2)}}$ [18.0$^\circ$ - 19.5$^\circ$]  & [170.0$^\circ$ - 180.0$^\circ$] & [180.0$^\circ$ - 190.0$^\circ$] \\
     \hline
     $\text{Bottom}\; _{\text{(layer3)}}$ [16.0$^\circ$ - 18.0$^\circ$]  & [168.0$^\circ$ - 180.0$^\circ$] & [180.0$^\circ$ - 192.0$^\circ$] \\
     \hline
\end{tabular}
\caption{$\theta-\phi$ regions used in study.}
\label{ta:region}
\end{table}
Table \ref{ta:t} presents the measured transmission values in the left and right regions of Mt. Vesuvius, evaluated across three different elevation layers, before and after the updated $\chi^2_{3p}$ cut criteria. As expected, the measured transmission increases with elevation angle, i.e. transmission is higher in the top layer compared to the bottom one, since the rock thickness decreases toward the summit of the volcano.
Across all angular bins, the ROSSO detector consistently yields higher transmission values than the NERO detector. In the top layer, the transmission on the right side is generally greater than on the left for all datasets, with the exception of NERO WP20. In contrast, this trend reverses in the middle and bottom layers, where the left-sided angular bins tend to show higher transmission values than the right-sided ones. Based on these transmission no obvious left-right asymmetry can be observed in the summit crate.

\begin{table}[h!]
\centering
\begin{tabular}{ m{2.6cm} | m{1cm} m{1cm} | m{1cm}  m{1cm} } 
\centering
  \multirow{2}{*}{Datasets} & \multicolumn{2}{c|}{No Cuts} & \multicolumn{2}{c}{New $\chi^2_{3p}$ cut} \\
  \cline{2-5}
   & $\text{T}_{left}$ & $\text{T}_{right}$ & $\text{T}_{left}$ & $\text{T}_{right}$ \\
   \hline
   \hline
  \multicolumn{5}{c}{Top} \\
  \hline
  ROSSO WP$15^{\circ}$C & 0.329 & 0.336 & 0.300 & 0.307 \\ 
  ROSSO WP$20^{\circ}$C & 0.344 & 0.348 & 0.308 & 0.315 \\ 
  NERO WP$15^{\circ}$C & 0.302 & 0.336 & 0.232 & 0.239 \\ 
  NERO WP$20^{\circ}$C & 0.284 & 0.348 & 0.259 & 0.259\\ 
  \hline
  \multicolumn{5}{c}{Middle} \\
  \hline
  ROSSO WP$15^{\circ}$C & 0.147 & 0.161 & 0.132 & 0.142 \\ 
  ROSSO WP$20^{\circ}$C & 0.168 & 0.160 & 0.144 & 0.139\\ 
  NERO WP$15^{\circ}$C & 0.153 & 0.149 & 0.118 & 0.112 \\ 
  NERO WP$20^{\circ}$C & 0.155 & 0.150 & 0.136 & 0.129 \\  
  \hline
  \multicolumn{5}{c}{Bottom} \\
  \hline
  ROSSO WP$15^{\circ}$C & 0.080 & 0.083 & 0.069 & 0.072 \\ 
  ROSSO WP$20^{\circ}$C & 0.087 & 0.090 & 0.074 & 0.076\\ 
  NERO WP$15^{\circ}$C & 0.072 & 0.075 & 0.055 & 0.058 \\ 
  NERO WP$20^{\circ}$C & 0.085 & 0.084 & 0.072 & 0.071\\ 
  \hline
\end{tabular}
\caption{Measured transmission in the left and right angular regions in three different layers, compared between original and $\chi^2_{3p}$ cut tracks selection criteria.}
\label{ta:t}
\end{table}

\section{Muon Flux Simulation}
Accurate simulations are essential for interpreting muographic data and for understanding the propagation of muons through the complex geological structure of Mt. Vesuvius. A detailed simulation workflow has been developed in MURAVES experiment to support data analysis. This includes the generation of cosmic ray showers using tools such as CORSIKA~\cite{CORSIKA}, CRY~\cite{CRY}, and EcoMug~\cite{EcoMug}, along with muon propagation studies through rock using PUMAS, TURTLE and MUSIC~\cite{MUSIC} Monte Carlo simulators. The detector response is simulated with Geant4~\cite{GEANT4}. The integration of these simulations allows for optimisation of reconstruction algorithms and better estimation of background.

In our previous simulation framework, we employed the TURTLE \cite{TURTLE} (Transport of Uncharged particles through matter) program in combination with PUMAS~\cite{PUMAS} (Parallel Ultra-fast MUon and tau neutrino Simulator) code to include the topography of the Mt. Vesuvius area and model muon transport through the mountain. The application of PUMAS and TURTLE was detailed in~\cite{2024}. Now a unified framework called MULDER (MUon simuLation for DEnsity Reconstruction) is adopted, which integrates both tools into a single utility library. MULDER is designed to compute local variations in atmospheric muon flux induced by geophysical features such as topography, using a Digital Elevation Model (DEM). This integration simplifies the simulation process, reduce potential sources of error and improving computational efficiency.

\begin{figure}[h!]
    \centering
    \begin{subfigure}[h!]{0.44\textwidth}
        \includegraphics[width=0.9\linewidth]{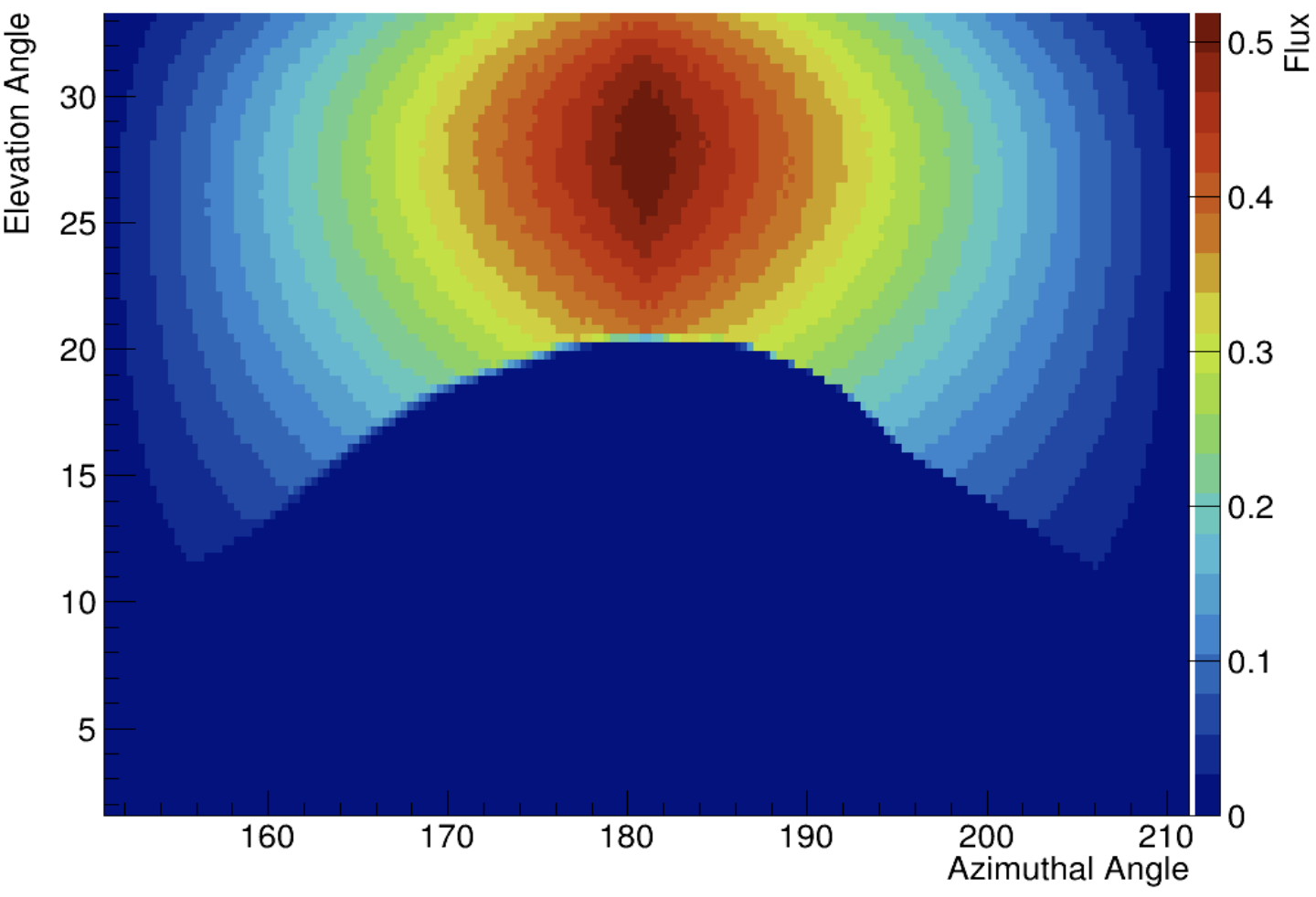}
        \caption{Simulated flux rate with Mulder library.}
        \label{fig:ROSSO}
    \end{subfigure}
    \hfill
     \begin{subfigure}[h!]{0.47\textwidth}
        \includegraphics[width=0.9\linewidth]{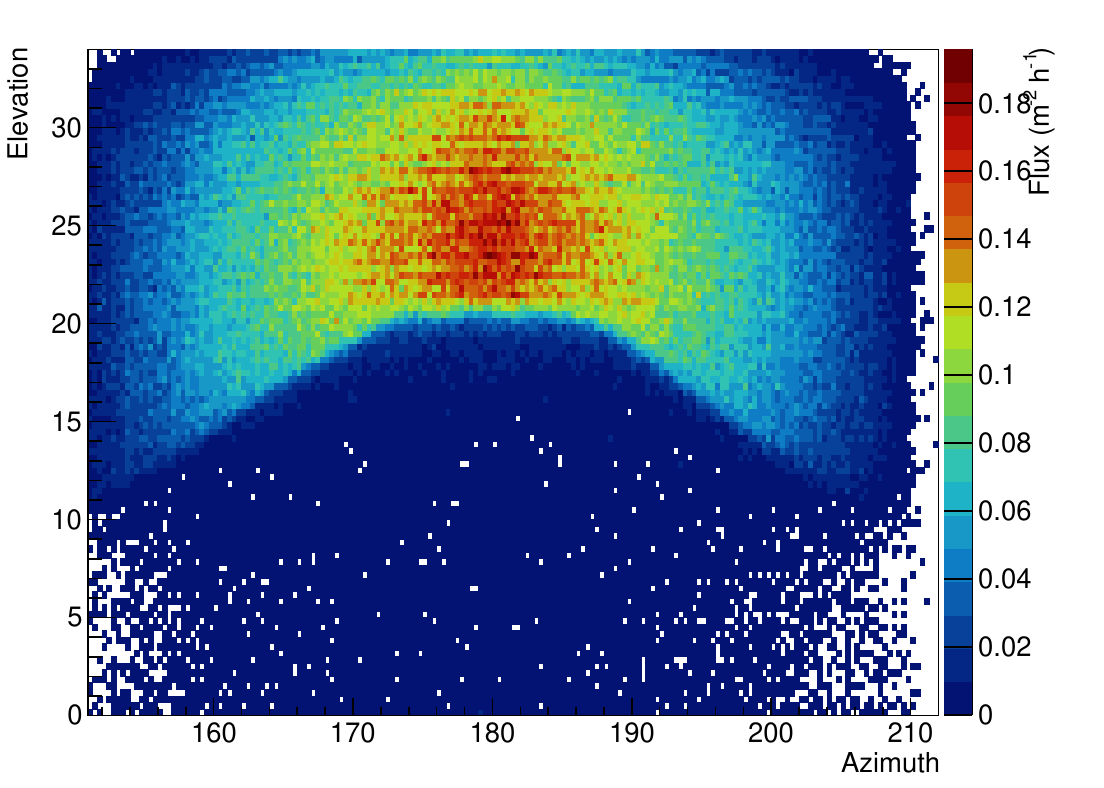}
         \caption{Experimental data of flux map of ROSSO detector collected over 1143.3 hours at WP $15^{\circ}$. }
         \label{fig:Siflux}
    \end{subfigure}
    \caption{An agreement is shown between simulated and experimental flux map.}
\end{figure}

The latest simulation results obtained with MULDER framework are shown in Fig. \ref{fig:Siflux}, while Fig. \ref{fig:ROSSO} presents the corresponding results for the ROSSO detector at a working point of $15^{\circ}$, based on 1143.3 hours of data collection. The improved agreement between simulated and experimental data validates the effectiveness of the "Golden Selection" criteria and use of MULDER framework, provides deeper insights into the internal structure of Mt. Vesuvius. Ongoing studies focus on detailed transmission comparisons and precise density extractions to enhance the accuracy of density reconstructions.  

\section{Conclusion}
The MURAVES experiment at Mt. Vesuvius has achieved advancements in both data analysis and simulation frameworks for muographic imaging. The implementation of the Golden Selection, a refined track quality criterion based on the $\chi^2$ value calculated from the first three detector planes, has led to a cleaner and more reliable sample of reconstructed muon tracks. This enhancement directly improves the resolution and accuracy of muographic images, enabling more precise density measurements of the volcanic structure.
In parallel, the deployment of the unified MULDER simulation framework has substantially improved our ability to model muon propagation through the complex topography of Mt. Vesuvius. The new framework provides a more streamlined and computationally efficient approach to simulating atmospheric muon fluxes.
Ongoing studies are focused on detailed transmission comparisons and precise density extraction techniques to further improve the accuracy of density reconstructions. These efforts aim to advance our understanding of the internal structure of Mt. Vesuvius summit crater and contribute to more reliable volcanic hazard assessment and risk mitigation strategies.

\section*{Conflict of Interest}
The authors have no conflicts to disclose.

\section*{Data Availability Statement}

The data that support the findings of this study are available from the corresponding author upon reasonable request.

\section*{References}
\nocite{*}
\bibliography{aipsamp.bib}% Produces the bibliography via BibTeX.

\end{document}